\documentstyle[11pt,moriond,epsfig]{article}

\def\mco{\multicolumn}
\def\spose#1{\hbox to 0pt{#1\hss}}
\def\lta{\mathrel{\spose{\lower 3pt\hbox{$\mathchar"218$}}
     \raise 2.0pt\hbox{$\mathchar"13C$}}}
\def\gta{\mathrel{\spose{\lower 3pt\hbox{$\mathchar"218$}}
     \raise 2.0pt\hbox{$\mathchar"13E$}}}

\def\etal {{\it et al\/}}

\def\plottwo#1#2{\centering \leavevmode
\epsfxsize= 0.49\columnwidth \epsfbox{#1}
\epsfxsize= 0.49\columnwidth \epsfbox{#2}}

\def\be{\begin{equation}}
\def\ee{\end{equation}}
\def\bea{\begin{eqnarray}}
\def\eea{\end{eqnarray}}
\def\lbl{\label}
\def\hm{{\cal H}^3}

\def\xic{\xi^{c}_\Phi({\bf x}, {\bf x}^\prime)}

\def\xiur{\xi^{u}_\Phi(r)}
\def\xiurj{\xi^{u}_\Phi(r_j)}

\def\taurec{\tau_{\rm rec}}

\begin{document}
{\hfill{\bf CITA-98-11}}
\vspace*{4cm}
\title{Probing Cosmic topology using CMB anisotropy}

\author{ Tarun Souradeep, Dmitry Pogosyan and  J. Richard Bond}

\address{Canadian Institute for Theoretical Astrophysics,\\ 
University of Toronto, ON M5S 3H8, Canada}

\maketitle\abstracts{ The measurements of CMB anisotropy have opened
up a window for probing the global topology of the universe on length
scales comparable to and beyond the Hubble radius. We have developed a
new method for calculating the CMB anisotropy in models with
nontrivial topology and apply it to open universe models with compact
spatial topology. We conduct a Bayesian probability analysis for a
selection of models which confronts the theoretical pixel-pixel
temperature correlation function with the {\sc cobe--dmr} data. Our
results demonstrate that strong constraints on compactness arise: if
the universe is small compared to the `horizon' size, correlations
appear in the maps that are irreconcilable with the observations.}

\noindent The remarkable degree of isotropy of the cosmic microwave
background (CMB) points to homogeneous and isotropic
Friedmann-Robertson-Walker (FRW) models for the universe. This
argument is a purely local one and does not refer to the global
topological structure of the universe. In fact, in the absence of
spatially inhomogeneous perturbations, a FRW model predicts an
isotropic CMB regardless of the global topological structure. However,
the observed large scale structure in the universe and CMB anisotropy
allude to the existence of small spatially inhomogeneous primordial
perturbations. The global topology of the universe does affect the
observable properties of the CMB anisotropy.  In compact universe
models, the finite spatial size usually implies a suppression of the
power in large scale perturbations and consequently the CMB anisotropy
is suppressed on angular scales above a characteristic angle related
to size of the universe. Another signature is the breaking of
statistical isotropy in characteristic patterns determined by the
photon geodesic structure of the compact manifold.

Much recent astrophysical data suggest the cosmological density
parameter, $\Omega_0$, is subcritical.~\cite{opencase} In the absence
of a cosmological constant, this would imply a hyperbolic spatial
geometry for the universe (commonly referred to as the `open' universe
in cosmological literature). The topologically trivial (simply
connected) hyperbolic 3-space, $\hm$, is non-compact and has infinite
size.  There are numerous theoretical motivations, however, to favor a
spatially compact universe.~\cite{cct_motive} To reconcile this with a
flat or hyperbolic geometry, consideration of models with non-trivial
topology is required. A compact cosmological model is constructed by
identifying points on the standard infinite flat or hyperbolic FRW
space under the action of a suitable discrete subgroup, $\Gamma$, of
the full isometry group, $G$, of the FRW space. The FRW spatial
hypersurface is the {\em universal cover}, tiled by copies of the
compact space, ${\cal M}$. Any point ${{\bf x}}$ of the compact space
has an image ${{\bf x}}_i = \gamma_i {{\bf x}}$ in each tile on the
universal cover, where $\gamma_i \in \Gamma$.

\begin{figure}[b]
\plottwo{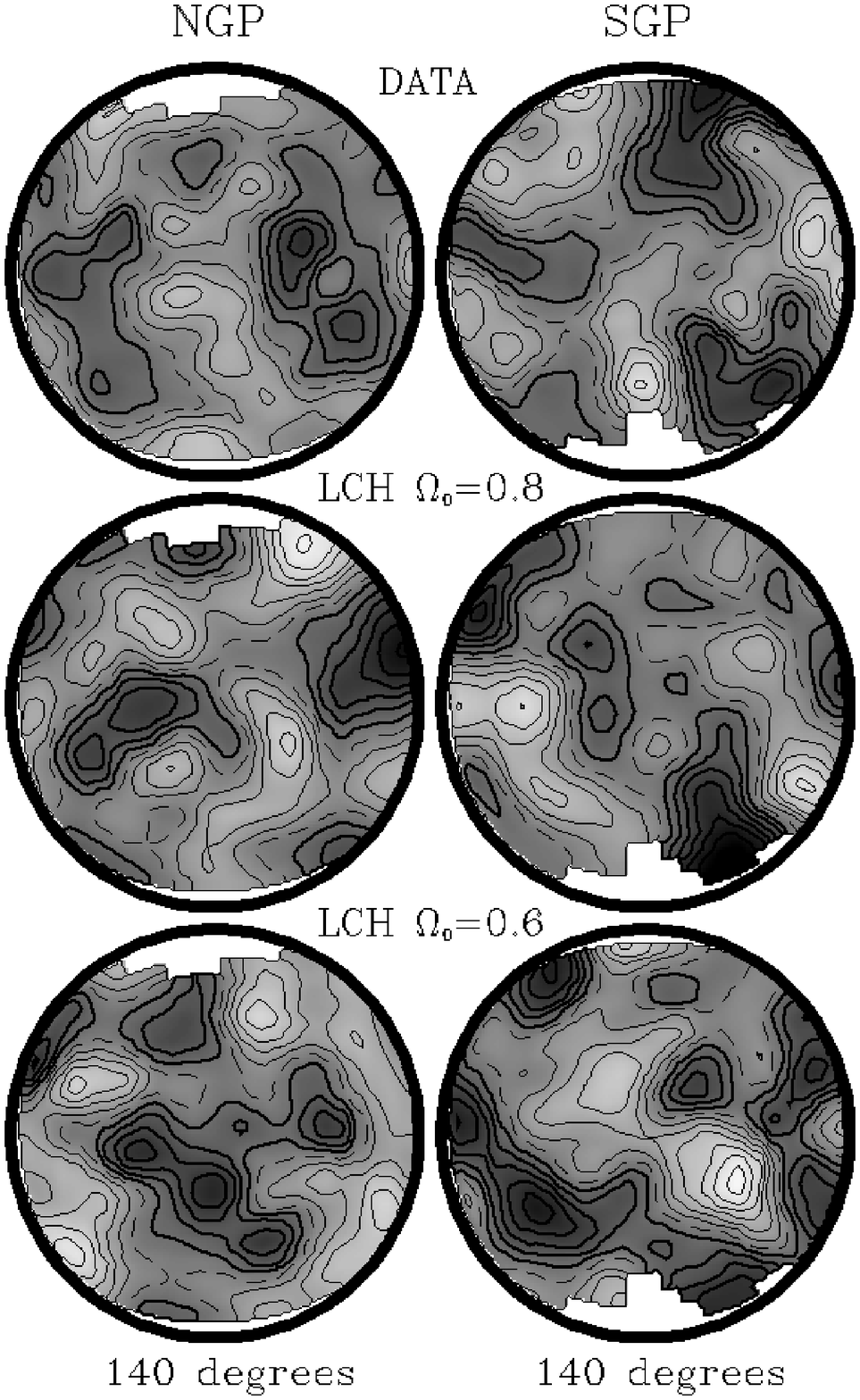}{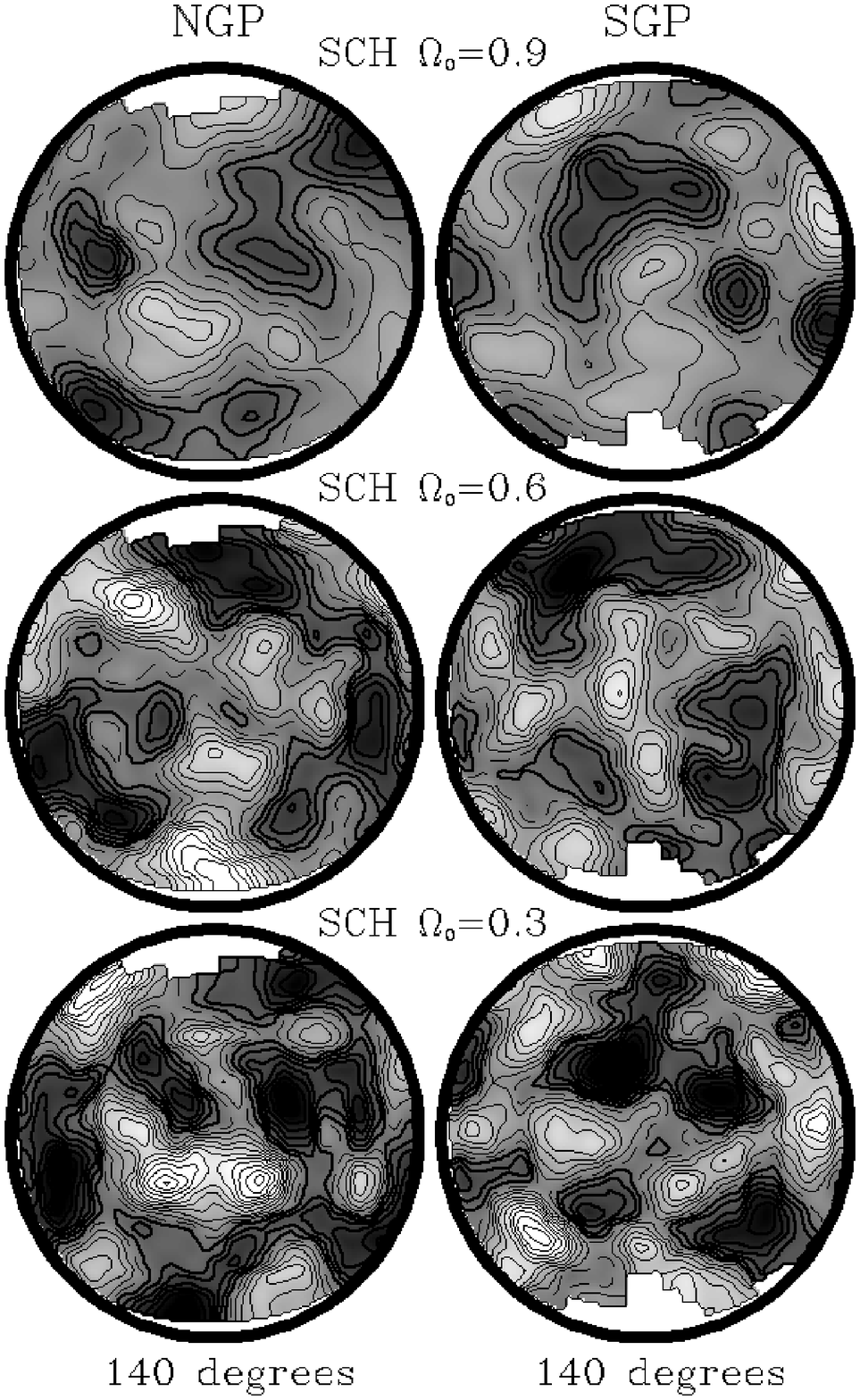}
\caption{The figure consists of two columns of CMB sky-maps showing a
pair of $140^\circ$ diameter hemispherical caps each, centered on the
South (SGP) and North (NGP) Galactic Poles, respectively. The map
labeled DATA, shows the {\sc cobe--dmr} 53+90+31 GHz A+B data after Wiener
filtering assuming a standard CDM model, normalized to {\sc cobe}.  The rest
of the five maps are one random realization of the CMB anisotropy in
two examples of compact hyperbolic (CH) spaces for several values of
$\Omega_0$ based on our theoretical calculations of $C(\hat q,\hat
q^\prime)$ convolved with the {\sc cobe--dmr} beam. 
Both surface and integrated (ISW) Sachs-Wolfe effects have been included in
$C(\hat q,\hat q^\prime)$. No noise was added. The
power was normalized to best match the {\sc cobe} data. The theoretical sky
was optimally filtered using the {\sc cobe} experimental noise to facilitate
comparison with data.  The maps labeled L(arge)CH refer to the CH
model $v3543(2,3)$. The right column shows the CMB maps for the
S(mall)CH model $m004(-5,1)$. (The model number associated with the
topology corresponds to that of the census of CH spaces in the
Geometry center, Univ. of Minnesota; SCH is one of the smallest and
LCH is one of the largest spaces in the census). LCH with
$\Omega_0=0.8$ is compatible with the data with a suitable choice of
orientation while all the others are ruled out (See
Table~\ref{table1}).  For all six maps, the average, dipole and
quadrupole of the $\vert b \vert > 20^\circ$ sky were also removed and a
$20^\circ$ Galactic latitude cut was used, with extra cuts to remove
known regions of Galactic emission proposed by the {\sc cobe} team
accounting for the ragged edges. The contours are linearly spaced at
$15~\mu {\rm K}$ steps.  The maps have been smoothed by a $1.66^\circ$
Gaussian filter.\hfill\mbox{}}
\label{fig1}
\end{figure}

For Gaussian perturbations, the angular correlation function, $C(\hat
q,\hat q^\prime)$, of the CMB temperature fluctuations in two
directions $\hat q$ and $\hat q^\prime$ in the sky completely encodes
the CMB anisotropy predictions of a model.  The dominant contribution
to the anisotropy in the CMB temperature measured with wide-angle beam
($\theta_{\sc fwhm} \gta 2^\circ~ \Omega_0^{1/2}$) comes from the
cosmological metric perturbations through the Sachs-Wolfe effect. The
angular correlation function of the CMB anisotropy, $C(\hat q,\hat
q^\prime)$, depends on the spatial two point correlation function,
$\xi_\Phi \equiv \langle\Phi({\bf x},\taurec)\Phi({\bf
x^\prime},\taurec)\rangle $ of the gravitational potential, $\Phi$, on
the hypersurface of last scattering.\footnote{ Other effects which
contribute to the CMB anisotropy at smaller angular scales can also be
approximated in terms of spatial correlation of quantities defined on
the hypersurface of last scattering.~\cite{us}} To calculate the
spatial correlation function on a compact hyperbolic (CH) manifold,
described by the corresponding $\Gamma$, we have developed a general
technique -- the {\em method of images}, which evades the difficult
problem
\footnote{The correlation function is usually computed using a mode
function expansion. However, obtaining closed form expressions for
eigenfunctions of the Laplacian may not be possible beyond the
simplest topologies and even numerical estimation is known to be
difficult in CH spaces.}  of solving for eigenfunctions of the
Laplacian on these manifolds.~\cite{us_texas} Using the method of
images, the spatial correlation function, $\xi^c_\Phi$, between two
points ${\bf x}$ and ${\bf x^\prime}$ on a compact space of volume
$V_{\cal M}$ can be expressed as~\cite{us}
\begin{equation}
\xic = \lim_{r_*\to\infty} \sum_{r_j < r_*} \xiurj - 
\frac{4\pi}{V_{\cal M}} \int_0^{r_*} dr ~\sinh^2r ~\xiur\,,~~~~~
r_j = d({\bf x}, \gamma_j {\bf x}^\prime)\, . 
\lbl{moi_final}
\end{equation}
a regularized sum over the correlation function, $\xi_\Phi^u$, on the
universal cover evaluated between ${\bf x}$ and images $\gamma_i{\bf
x^\prime}$ of ${\bf x^\prime}$.  Numerically it suffices to evaluate
the above expression up to $r_*$ a few times the curvature radius,
$d_c$, to reach a convergent result. We then integrate $\xi^c_\Phi$ along
photon trajectories to get $C(\hat q,\hat q^\prime)$ which
includes both surface and integrated (ISW) Sachs-Wolfe effects.

The CMB photons can be viewed as propagating to the observer from a
$2$-sphere of radius, $R_H$, -- the sphere of last scattering
(SLS). In contrast to the topologically trivial models, widely
separated pixels in compact spaces can still have strong correlations
if in the sum over images, eq.~(\ref{moi_final}), one of the images of
${\bf x^\prime}$ happens to be close to ${\bf x}$. The dependence of
the spatial correlations on the anisotropic distribution of images
leads to a characteristic statistical anisotropy in the CMB in compact
models. These effects are pronounced when the compact space fits well
within the SLS, but persist at an observable level even when $V_{\cal
M}$ is comparable to (or somewhat bigger than) $V_{\sc sls}$, the
volume of SLS.  If SLS does not fit completely inside a single tile --
a copy of the compact space, the CMB temperature values will be
identical along pairs of circles if temperature fluctuations are
dominated by the surface terms at the SLS.~\cite{circles} This pattern
of matched circles is one specific manifestation of the angular
patterns in $C(\hat q,\hat q^\prime)$.

Figure~\ref{fig1} compares theoretical realizations of the CMB
anisotropy in a selection of CH models with the {\sc cobe--dmr} data.
In this work, the six {\sc cobe--dmr} four-year maps~\cite{dmr4} are
first compressed into a (A+B)(31+53+90 GHz) weighted-sum map, with the
customized Galactic cut advocated by the {\sc dmr} team. There is no
effective loss of information when we do further data compression by
using $5.2^\circ\times5.2^\circ$ pixels.~\cite{bdmr294} The
theory and data maps have been postprocessed so as to facilitate a
fair visual comparison. The incompatibility of models with small
$V_{\cal M}/V_{\sc sls}$ (SCH-$\Omega_0=0.3,0.6$) is visually obvious:
the best fit amplitudes are high which is reflected in the steeper hot
and cold features. Although, SCH-$\Omega_0=0.9$ and LCH-$\Omega_0=0.6$
do not appear grossly inconsistent, the intrinsic anisotropic
correlation pattern is at odds with the data.

We have carried out a full Bayesian analysis of the probability of the
CH models given the {\sc cobe--dmr} 4yr data. In Table~\ref{table1} we
present the {\em relative likelihood} of the selected models to that
of the infinite, $\hm$, model with the same $\Omega_0$. (The {\sc
cobe} data alone does not strongly differentiate between the infinite
hyperbolic models with different $\Omega_0$.) The anisotropy of the
theoretical $C(\hat q,\hat q^\prime)$ causes the likelihood of compact
models to vary significantly with the orientation of the space with
respect to the sky, depending on how closely the features in the
single data realization available match (or mismatch) the pattern in
$C(\hat q,\hat q^\prime)$. Some optimal orientations may also have the
``ugly'' correlation features hidden in the Galactic cut.  We analyzed
$24$ different orientations for each of our models and found that only
the model with $V_{\cal M} > V_{\sc sls}$ (LCH-$\Omega_0=0.8$) cannot
be excluded (at one orientation this model is even preferable to
standard CDM; this raises a question of the statistical significance
of any detection of intrinsic anisotropy of a space when only a single
realization of data is available).

Similar conclusions were reached by some of the authors (JRB, DP and
I. Sokolov~\cite{us_torus}) for flat toroidal models.  Comparison of
the full angular correlation with {\sc cobe} data led to a much
stronger limit on the compactness of the universe than limits from
other methods.~\cite{tor_refs} The main result of the analysis was
that $V_{\sc sls}/V_{\!\cal M} < 0.4$ at $95\%~CL$ for the equal-sided
$3$-torus. For non compact $1$-torus, the constraint on the most
compact dimension is not quite as strong.

In summary, our results demonstrate that the {\sc cobe} data can put strong
constraints on the compact models of the universe. If the universe is
small compared to the `horizon' size, correlations appear in the maps
that are irreconcilable with the large angle {\sc cobe--dmr} data.
\vspace*{-2mm}
\begin{table}[htb]
\caption{The Log-likelihoods of the compact hyperbolic models relative
to the infinite models with same $\Omega_0$ are listed below.  The
likelihoods are calculated by comparison with {\sc cobe--dmr} data. The
three columns of Log-likelihood ratios correspond to the best, second
best and worst values that we have obtained amongst $24$ different
rotations of the compact space relative to the sky. The number in
brackets gives a convenient, albeit crude, translation to gaussian likelihood.
Only the last model can be reconciled
with the {\sc cobe--dmr} data.\hfill\mbox{}\label{logprob_tab}}
\vspace{0.3cm}
\begin{center}
\begin{tabular}{|c|c|c|c|c|c|}
\hline Topology &$\Omega_0$&$V_{\sc sls}/V_{\!\cal
M}$&\mco{3}{|c|}{Relative Log. Likelihood (Gaussian approx.)}\\
\cline{4-6} & & &\mco{3}{|c|}{Orientation}\\
&    &    &`best'&`second best'&`worst'\\
\hline
$m004(-5,1)$ &0.3&153.4 & -35.5 (8.4$\sigma$)& -35.7 (8.4$\sigma$) 
& -57.9 (10.8$\sigma$)\\
&0.6&19.3 & -22.9 (6.8$\sigma$)& -23.3 (6.8$\sigma$) 
& -49.4 ( 9.9$\sigma$)\\
$V_{\cal M}/d_c^3=0.98$      &0.9&1.2  & -4.4  (3.0$\sigma$)& -8.5  (4.1$\sigma$) 
& -37.4 ( 8.6$\sigma$)\\
\hline
$v3543(2,3)$ & 0.6 &2.9 & -3.6 (2.7$\sigma$) 
& -5.6 (3.3$\sigma$)& -31.0 (7.9$\sigma$)  \\
$V_{\cal M}/d_c^3=6.45$ & 0.8 &0.6 &  2.5 (2.2$\sigma$) 
& -0.8 (1.3$\sigma$)   & -12.6 (5.0$\sigma$)  \\
\hline
\end{tabular}
\end{center}
\lbl{table1}
\end{table}

\end{document}